\documentclass[preprint,12pt, a4paper]{elsarticle}

\usepackage{amssymb}
\usepackage{amsmath}
\usepackage{bm}
\usepackage{lineno}
\usepackage{caption}
\usepackage{subcaption}
\usepackage{tikz}
\usepackage[colorlinks=true,citecolor=blue,linkcolor=blue,urlcolor=blue]{hyperref}
\usepackage[frozencache]{minted}
\setminted{fontsize=\small}
\usepackage{float}
\restylefloat{table}
\journal{SoftwareX}
\usepackage[draft]{fixme}
\fxsetup{
  layout=marginnote,
  marginface=\normalfont\tiny,
  envface=,
  inlineface=,
  theme=color
}
\setcitestyle{maxnames=6}

\usepackage[mathletters]{ucs} 
\usepackage[utf8x]{inputenc}
\usepackage{xcolor}

\begin{document}

\begin{frontmatter}


%\title{SpinGlassPEPS.jl: low-energy spectrum for low-dimensional Ising models}

\title{SpinGlassPEPS.jl: \\ Tensor-network package for Ising-like optimization on quasi-two-dimensional graphs}

\author[1]{Tomasz \'{S}mierzchalski}
\author[2,3]{Anna M. Dziubyna}
\author[1]{Konrad Ja\l{}owiecki}
\author[1]{Zakaria Mzaouali}
\author[1]{{\L}ukasz Pawela}
\author[1]{Bart\l{}omiej Gardas}
\author[2]{Marek M. Rams}

\address[1]{Institute of Theoretical and Applied Informatics, Polish Academy of Sciences, Ba{\l}tycka 5, 44-100 Gliwice, Poland}
\address[2]{Jagiellonian University, Institute of Theoretical Physics, \L{}ojasiewicza 11, 30-348 Krak\'ow, Poland}
\address[3]{Jagiellonian University, Doctoral School of Exact and Natural Sciences, \L{}ojasiewicza 11, 30-348, Krak\'ow, Poland}

\begin{abstract}
This work introduces \texttt{SpinGlassPEPS.jl}, a software package implemented
in Julia, designed to find low-energy configurations of generalized Potts
models, including Ising and QUBO problems, utilizing heuristic tensor network
contraction algorithms on quasi-2D geometries. In particular, the package
employs the Projected Entangled-Pairs States to approximate the Boltzmann
distribution corresponding to the model's cost function. This enables an
efficient branch-and-bound search (within the probability space) that exploits
the locality of the underlying problem's topology. As a result, our software
enables the discovery of low-energy configurations for problems on quasi-2D
graphs, particularly those relevant to modern quantum annealing devices. The
modular architecture of \texttt{SpinGlassPEPS.jl} supports various contraction
schemes and hardware acceleration. 
\end{abstract}

\end{frontmatter}

\section*{Required Metadata}

\begin{table}[H]
\begin{tabular}{|l|p{6.5cm}|p{6.5cm}|}
\hline
\textbf{Nr.} & \textbf{Code metadata description} & \textbf{Please fill in this column} \\
\hline
C1 & Current code version & v1.4.1\\
\hline
C2 & Permanent link to code/repository used for this code version & \url{https://github.com/euro-hpc-pl/SpinGlassPEPS.jl} \\
\hline
C4 & Legal Code License   & Apache 2.0 \\
\hline
C5 & Code versioning system used & git \\
\hline
C6 & Software code languages, tools, and services used & Julia \\
\hline
C7 & Compilation requirements, operating environments \& dependencies & \texttt{Julia 1.11, CUDA.jl v5, TensorOperations.jl, Memoize.jl} \\
\hline
C8 & If available, Link to developer documentation/manual & \url{https://euro-hpc-pl.github.io/SpinGlassPEPS.jl/dev/}\\
\hline
C9 & Support email for questions & \verb|lpawela@iitis.pl|\\
\hline
\end{tabular}
\caption{Code metadata}
\label{} 
\end{table}

% \linenumbers
\section{Motivation and significance}

\texttt{SpinGlassPEPS.jl} provides a robust software package designed to find
low-energy configurations for generalized Potts models~\cite{Kinzel1981},
including Ising~\cite{Lucas2014} or equivalently Quadratic Unconstrained Binary
Optimization (QUBO) problems. By leveraging heuristic tensor network algorithms,
specifically Projected Entangled-Pairs States
(PEPS)~\cite{nishino_two-dimensional_2001,
verstraete_criticality_2006,verstraete_matrix_2008,Orus2014}, the package
enables efficient exploration of low-energy solutions within complex problem
topologies. This capability is particularly significant in the context of
current quantum and classical annealing
devices~\cite{King_quantum_2023,GotoSBM,Wang2023}, where efficient and scalable
solutions are essential.

The significance of \texttt{SpinGlassPEPS.jl} lies in its ability to lower the
entry barrier to applying advanced tensor network methods in classical
optimization. It offers a modular architecture that supports various contraction
schemes and hardware acceleration, making it adaptable to a wide range of
problem instances. The software has been extensively benchmarked~\cite{main} on
problems defined on D-Wave's Pegasus and Zephyr
geometries~\cite{Boothby2021,Dattani2019}.

\section{Software description}
\begin{figure}
        \centering
\includegraphics[width=0.99\textwidth]{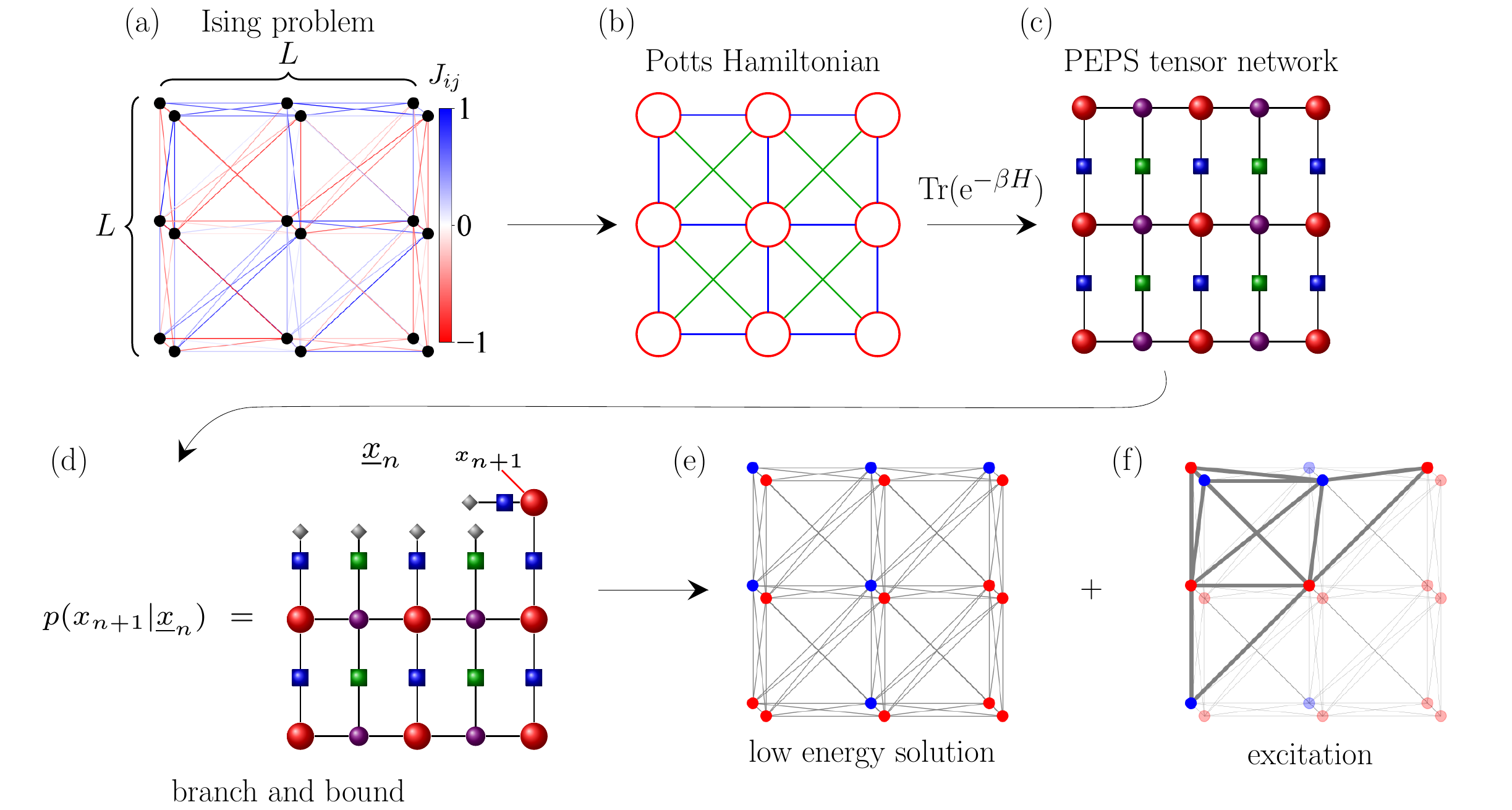}
        \caption{Execution flow. The Ising problem in (a) is mapped to a Potts
        Hamiltonian defined on a king's graph in (b). This allows the partition
        function of that Hamiltonian to be represented as a PEPS tensor network
        on a square lattice, as in (c). The main algorithm executes the branch
        and bound search in the probability space, building the most probable
        configurations by adding one Potts variable at a time. The marginal
        conditional probabilities follow from an approximate contraction of the
        corresponding tensor network in (d). The full branch and bound sweep
        results in a candidate for the  most probable (ground state)
        configuration in (e), together with a set of localized excitations on
        top of it in (f).}
        \label{fig:1}
    \end{figure}

\texttt{SpinGlassPEPS.jl} is a collection of Julia~\cite{Bezanson2017} packages
implementing heuristic tensor-network based algorithm to find low-energy states
and their corresponding energies (i.e., the spectrum) of generalized Potts
model,
\begin{equation}
    E(\bm{x}) = \sum_{\langle m, n\rangle \in \mathcal{F}} E_{m,n}(x_m, x_n) + \sum_{n \in \mathcal{W}} E_n(x_n),
\label{eq:Potts}
\end{equation}
defined on a graph $\mathcal{G} = (\mathcal{W}, \mathcal{F})$ specified by its
edges, $\mathcal{F}$, and vertices, $\mathcal{W}$. The method tackles a family
of sparse two-dimensional graphs called king's graphs~\cite{Chang2013}, see
Fig.~\ref{fig:1}(b). A particular problem instance is defined by real-valued
functions, $E_n(x_n)$ and $E_{m,n}(x_m, x_n)$, where $n$, $m \in \mathcal{W}$.

In particular, this includes the Ising model,
\begin{equation}
    E(\bm{s}) =  \sum_{\langle i, j\rangle \in \mathcal{E}} J_{ij} s_i s_j + \sum_{i \in \mathcal{V}} h_i s_i,
\label{eq:Ising}
\end{equation}
defined on $\mathcal{G}' = (\mathcal{V}, \mathcal{E})$, where $i, j \in
\mathcal{V}$, $J_{ij}, h_i \in \mathbb{R}$ and $s_i \in \{ -1, 1\}$. The Ising
model can be considered a special case of the Potts model, and we extend this by
allowing clusters of spin variables to be grouped into effective Potts degrees
of freedom with higher dimensions. This approach enables the package to manage
more complex quasi-2D graph geometries, including those relevant to current
quantum and classical annealing device architectures.

The algorithm executes a branch-and-bound search~\cite{Morrison2016} within the
probability space defined by the Boltzmann distribution at a specific inverse
temperature $\beta$. This search process involves constructing a predefined
number of the most probable local configurations of Potts variables, progressing
sequentially from one vertex to the next until the entire system has been
explored. The emphasis on king's graphs is incorporated into the algorithm at
two distinct levels.

First, in calculating marginal probabilities, we employ a tensor network
representation of the classical Boltzmann distribution. For two-dimensional
systems, this representation takes the form of PEPS. The marginal and
conditional probabilities are derived from the contraction of the tensor
network. Although the problem of contracting these networks is formally
\#P-hard~\cite{Schuch2007}, we utilize an established approximate (heuristic)
framework to contract two-dimensional
PEPS~\cite{verstraete_matrix_2008,Orus2014}.

Furthermore, we leverage the locality of interactions within the graph to expand
the search space by merging partial trial configurations that are equivalent in
terms of the marginal conditional probabilities considered by the algorithm.
Specifically, partial configurations with identical values of Potts variables on
the boundary adjacent to unexplored regions of the graph are combined. This
approach also enables the identification of information regarding local
excitations in the system. When two partial configurations share boundary
variables, they have a well-defined energy associated with the bulk variables,
where one configuration represents a local ground state and the other an
excitation localized in the bulk. By collecting such information throughout the
search, the algorithm generates a candidate for the ground state along with a
set of excited low-energy states, effectively characterizing the low-energy
manifold of the optimization problem.

The algorithm workflow is outlined in Fig.~\ref{fig:1}. The package is designed
with a modular architecture that captures the relationships between the
high-level concepts employed by the algorithm. This modularity enables the
integration of various contraction schemes, leveraging the internal structures
of individual tensors within the network, and allows for hardware acceleration,
all facilitated by the multiple dispatch capabilities of the Julia language.
Detailed explanations of the algorithm's mechanics, along with extensive
benchmarks for Pegasus and Zephyr geometries, relevant for D-Wave quantum
annealers, are provided in~\cite{main}.

\subsection{Software Architecture}
\label{sec:architecture}

\begin{figure}[h!]
    \centering
    \includegraphics[width=\textwidth]{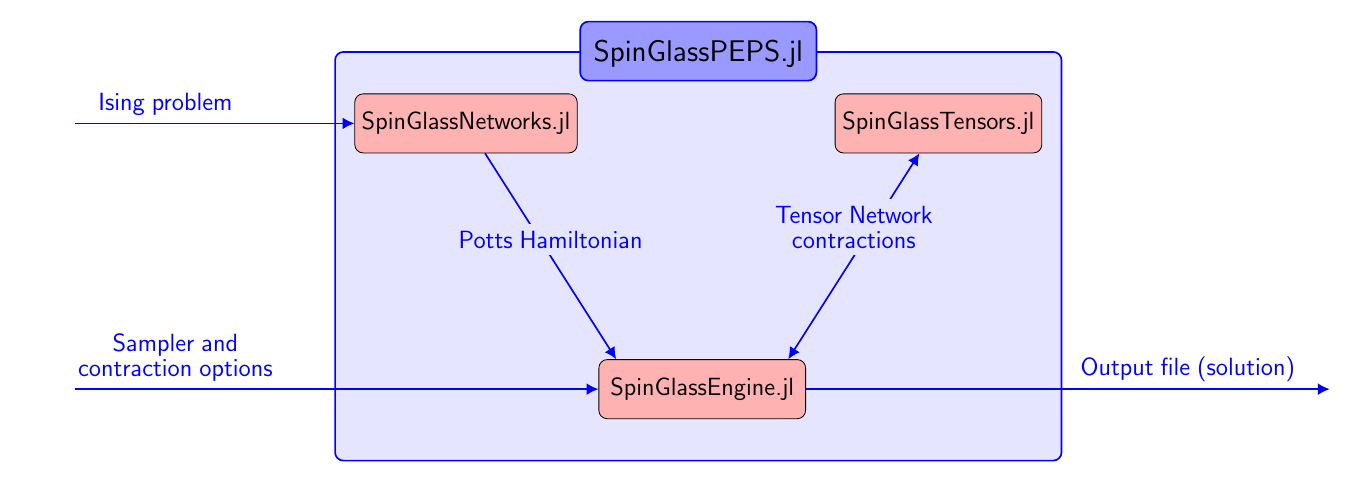}
    \caption{Interoperability between all \texttt{SpinGlassPEPS.jl} packages.
    The input Ising problem file is processed by~\texttt{SpinGlassNetworks.jl},
    transforming it into a Potts Hamiltonian. The latter is then passed
    to~\texttt{SpinGlassEngine.jl}, together with solver's and contraction's
    parameters. The~\texttt{SpinGlassEngine.jl} module serves as the core branch
    and bound solver. It passes the problem of marginal probabilities'
    calculation to \texttt{SpinGlassTensors.jl}, that constructs and
    approximately contracts the corresponding tensor network. Finally, the
    solution (ground state, excitations, and their energies) is returned to the
    user as an output. }
    \label{fig:2}
\end{figure}

\texttt{SpinGlassPEPS.jl} is composed of three independent sub-packages, each of
which is responsible for distinct elements of the workflow, see
Fig.~\ref{fig:2}. Namely,

\begin{itemize}
    \item \texttt{SpinGlassEngine.jl} serves as the core package, consisting of
    routines for executing the branch-and-bound method (with the ability to
    leverage the problem's locality) for a given Potts instance. It also
    includes capabilities for reconstructing the low-energy spectrum from
    identified localized excitations and provides a tensor network constructor.

    \item \texttt{SpinGlassNetworks.jl} facilitates the generation of an Ising
    graph from a given instance using a set of standard inputs (e.g., instances
    compatible with the Ocean environment provided by D-Wave) and supports
    clustering to create effective Potts Hamiltonians.

    \item \texttt{SpinGlassTensors.jl} offers essential tools to create and
    manipulate tensors that build the PEPS network, with support for CPU and GPU
    utilization. It manages core operations on tensor networks, including
    contraction, using the boundary Matrix Product State
    approach~\cite{verstraete_matrix_2008}. This package primarily functions as
    a backend and users generally do not interact with it directly.
    
\end{itemize}
    The \texttt{SpinGlassPEPS.jl} can be installed using the Julia package
    manager for Julia v1.11. In Julia \texttt{REPL} type \texttt{]} to enter
    \texttt{Pkg REPL}, then type: \mint{Julia}|pkg> add SpinGlassPEPS|

\subsection{Software Functionalities}
For Ising/QUBO problems, the package requires instances to be provided in a
specific format, similar to the one used in Stanford Gset (reduced from
Max-Cut)~\cite{Gset}. In this format, the spins are numbered $1$ to $N$ and
arranged in rows as $i$ $j$ $v$, where $v$ represents the coupling value
between vertices $i$ and $j$, or the local magnetic field when $i=j$. 
   
The algorithm produces a low-energy spectrum as its output, detailing the states
and their corresponding energies for the given input instance. In addition, the
package provides supplementary information, such as the estimated probabilities
of these states derived from the approximate contraction of the tensor network.
    
The package's modular structure offers a range of options to control various
aspects of the main algorithm. These include the details and control parameters
of the tensor network contraction schemes, the ability to utilize either CPU or
GPU for low-level operations, and the choice between constructing the tensor
network in a dense or sparse format. The sparse format leverages the internal
structures of individual tensors and is particularly essential for handling
large clusters, such as those relevant to Pegasus and Zephyr graphs. Detailed
information on all required and optional parameters is available in the
documentation~\cite{documentation}.
    
\section{Illustrative Examples}
We illustrate the capabilities and modularity of \texttt{SpinGlassPEPS.jl} by
addressing two distinct problems. First, we solve an Ising problem defined on a
quasi-2D graph, see Fig.~\ref{fig:1}(a). Although this example focuses on a
specific topology, the approach is applicable to other structures, including
those used in D-Wave systems. The second example tackles an inpainting problem
formulated through the Potts model~\cite{Lellmann2011, Kappes2013, openGM}.

The code for all the examples presented below can be found in the main
repository~\cite{ourGit} folder ``examples''. To run them, first, clone the
repository. Then, run Julia inside the ``examples'' folder. The environment
needed to execute the provided code is given by the \texttt{Project.toml} file.
It can be activated by typing \texttt{]} in Julia \texttt{REPL} and then:

\mint{Julia}|pkg> activate .|
\mint{Julia}|pkg> instantiate|

\noindent next in Julia \texttt{REPL} type:

\mint{Julia}|julia> include("ising_model_on_a_kings_graph.jl")|

\noindent Alternatively, one can run the examples by typing: 

\mint{bash}|$ julia --project=. -e "using Pkg; Pkg.instantiate()"|
\mint{bash}|$ julia --project=. ising_model_on_a_kings_graph.jl|
\noindent in the preferred shell.

While the Ising model on a king's graph should compute relatively fast, the
latter two examples (an inpainting example and a small pegasus-type graph) can
take a long time to finish if run on the CPU only. It is recommended to run them
using GPU.

\subsection{Ising model on a king's graph} 

In the listing below, we show a complete Julia script to define and solve an
Ising problem defined on a graph in Fig.~\ref{fig:1}(a). 

\begin{minted}[escapeinside=||]{Julia}
using SpinGlassPEPS

function get_instance(topology::NTuple{3, Int})
    m, n, t = topology
    "$(@__DIR__)/instances/$(m)x$(n)x$(t).txt"
end

function run_square_diag_bench(::Type{T}; topology::NTuple{3, Int}) where {T}
    m, n, _ = topology
    instance = get_instance(topology)
    lattice = super_square_lattice(topology)

    hamming_dist = 5
    eng = 10

    best_energies = T[]

    potts_h = potts_hamiltonian(
        ising_graph(instance),
        spectrum = full_spectrum,
        cluster_assignment_rule = lattice,
    )

    params = MpsParameters{T}(; bond_dim = 16, num_sweeps = 1)
    search_params = SearchParameters(; max_states = 2^8, cut_off_prob = 1E-4)

    for transform ∈ all_lattice_transformations
        net = PEPSNetwork{KingSingleNode{GaugesEnergy}, Dense, T}(
            m, n, potts_h, transform,
        )

        ctr = MpsContractor(SVDTruncate, net, params; 
            onGPU = false, beta = T(2), graduate_truncation = true,
        )

        single = SingleLayerDroplets(eng, hamming_dist, :hamming)
        merge_strategy = merge_branches(
            ctr; merge_type = :nofit, update_droplets = single,
        )

        sol, _ = low_energy_spectrum(ctr, search_params, merge_strategy)

        push!(best_energies, sol.energies[1])
        clear_memoize_cache()
    end

    ground = best_energies[1]
    @assert all(ground .≈ best_energies)

    println("Best energy found: $(ground)")
end

T = Float64
@time run_square_diag_bench(T; topology = (3, 3, 2))
\end{minted}

In the example above, \mintinline{Julia}{transform} specifies a rotation of the
quasi-2D graph. The branch-and-bound search operates with a fixed order of
sweeping through local variables, corresponding to the fixed order of tensor
network contraction. By rotating the network, the search and contraction can
effectively begin from different starting points on the 2D grid, thereby
enhancing the stability of the results. This example performs the search across
all eight possible transformations of the 2D grid, comparing the best energies
obtained for each configuration.

Other control parameters include \mintinline{Julia}{Sparsity}, which determines
whether dense or sparse tensors should be used. The concept of sparse tensors
was introduced to manage large clusters containing $\mathcal{O}(10{-}20)$ Ising
variables (\emph{i.e.}, spins) each, where explicit construction of PEPS
tensors, triggered by \mintinline{Julia}{Sparsity=Dense}, quickly becomes
infeasible. Conversely, \mintinline{Julia}{Sparsity=Sparse} circumvents the need
for direct construction of individual tensors by performing optimal contractions
on small tensor diagrams utilizing internal structure of individual tensors.
These diagrams are then combined to efficiently contract the entire network.
Finally, \mintinline{Julia}{Node = KingSingleNode{GaugesEnergy}} specifies the
type of the node used within the tensor networks (e.g., switching between king's
graph or a square lattice); \mintinline{Julia}{Layout = GaugesEnergy} denotes
the division of the PEPS network into boundary Matrix Product
States~\cite{verstraete_matrix_2008,Orus2014} used to contract the network. 

Low-energy excitations (\emph{i.e.}, droplets) above the best solution can be
identified during the optional \mintinline{Julia}{merge_branches} step. This
option can be provided as an argument to the function that executes the
branch-and-bound algorithm, \mintinline{Julia}{low_energy_spectrum}
(see~\cite{main} for extended discussion). To view all droplets found, one may
invoke \mintinline{Julia}{unpack_droplets(solution)}.

The algorithm searches for diverse excitations within a specified energy range
above the ground state. An excitation is accepted only if its Hamming distance
from any previously identified excitation exceeds a predefined threshold. This
is governed by the parameters \mintinline{julia}{energy_cutoff}, which sets the
maximum allowed energy above the ground state, and
\mintinline{julia}{hamming_cutoff}, which determines the minimum Hamming
distance required between excitations for them to be considered distinct.

\begin{figure}
    \centering
    \includegraphics[width=\textwidth]{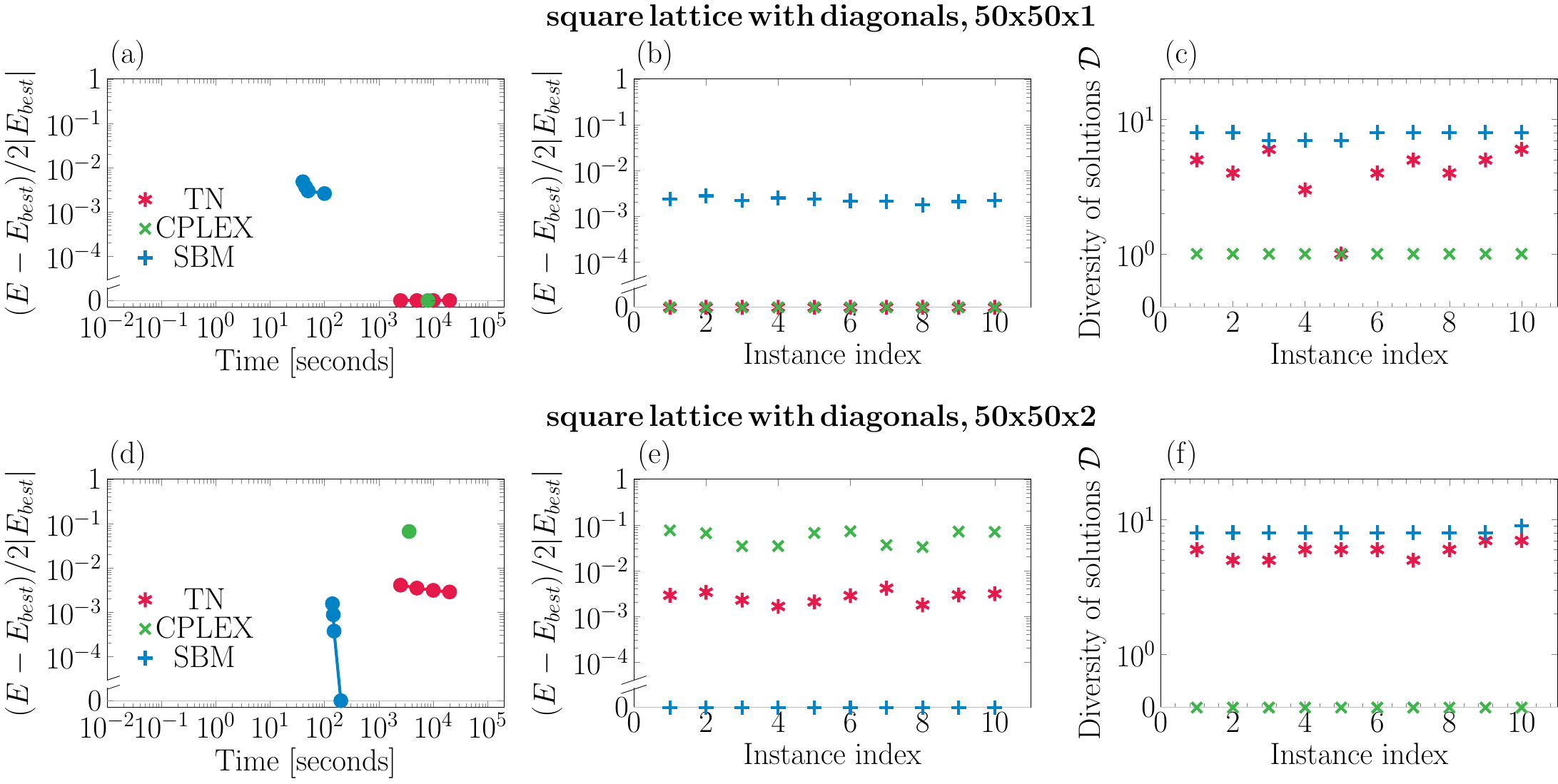}
    \caption{Benchmarking \texttt{SpinGlassPEPS.jl} for two sets of Ising
    problems defined on graphs from Fig.~\ref{fig:1}(a) with $N = 50 \times 50
    \times d$ spins, for $d = 1$ (top row) and $d=2$ (bottom row). As reference
    solvers, We employ a Simulated Bifurcation machine (SBM)~\cite{Goto2019} and
    CPLEX. Panels (a) and (d) show time to solution to the best energy for a
    median instance ($E_{\rm best}$ is the best results among the considered
    solvers), and in (b) and (e), we show instance-wise results for $10$
    instances. In (c) and (f), w2e show the diversity of obtained solutions,
    \emph{i.e.}, the number of solutions within approximation ratio $a_r = 0.01$
    ($E - E_{\rm best} < a_r \cdot 2 \cdot E_{\rm best}$), where each pair has a
    Hamming distance greater than $N / 2$.  
    }
    \label{fig:benchmark}
\end{figure}

We present a selected set of benchmark results in Fig.~\ref{fig:benchmark},
focusing on two problem sets with $2500$ and $5000$ spins (a $50 \times 50$ grid
with $1$ and $2$ spins per cluster, respectively). The reference results are
compared against those obtained from Simulated Bifurcation Machine
(SBM)~\cite{GotoSBM} and CPLEX solvers~\cite{CPLEX}.

For an instances with $2500$ spins, \texttt{SpinGlassPEPS.jl} successfully found
the ground states for all cases, as certified by CPLEX, outperforming SBM in
these tests. This demonstrates that, in certain scenarios (e.g., king's graph),
our approach can exceed state-of-the-art methods. For $5000$ spins, SBM delivers
the best results; however, \texttt{SpinGlassPEPS.jl} still performs better than
CPLEX, which faces challenges with larger problem sizes.

Both \texttt{SpinGlassPEPS.jl} and SBM can output multiple solutions
(\emph{i.e.}, low-energy states), whereas CPLEX generates only one. When
considering time to solution, our method is typically the most time-intensive,
while SBM is the fastest across all tested solvers since it can be
GPU-accelerated efficiently.

\subsubsection{Large unit cells}

\begin{figure}[h]
    \begin{subfigure}[b]{0.3\textwidth}
        \centering
        \includegraphics[width=\textwidth]{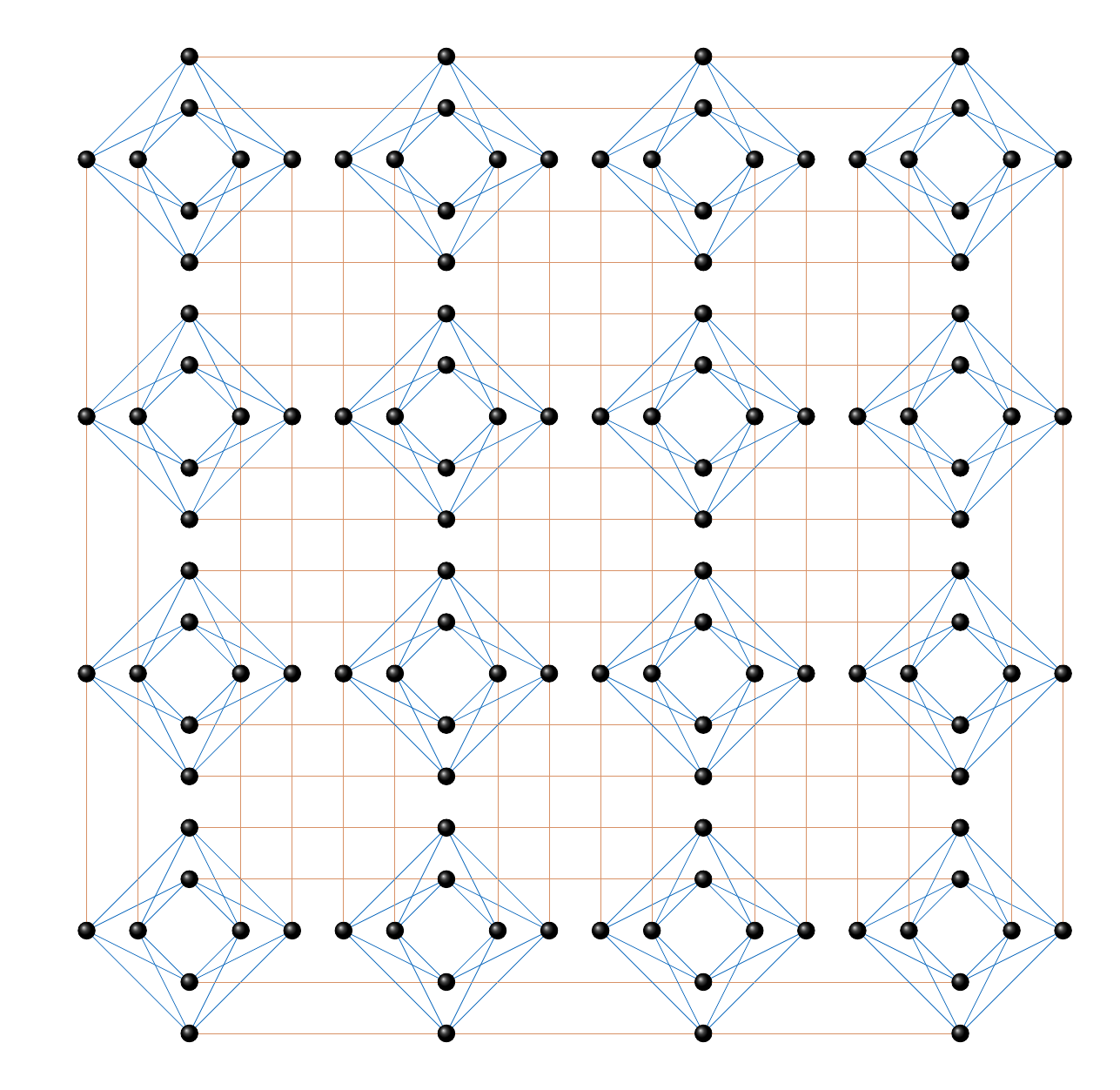}
         \caption{Chimera}
         \label{fig:chimera-qpu}
    \end{subfigure}
    \hfill
    \begin{subfigure}[b]{0.3\textwidth}
        \centering
        \includegraphics[width=\textwidth]{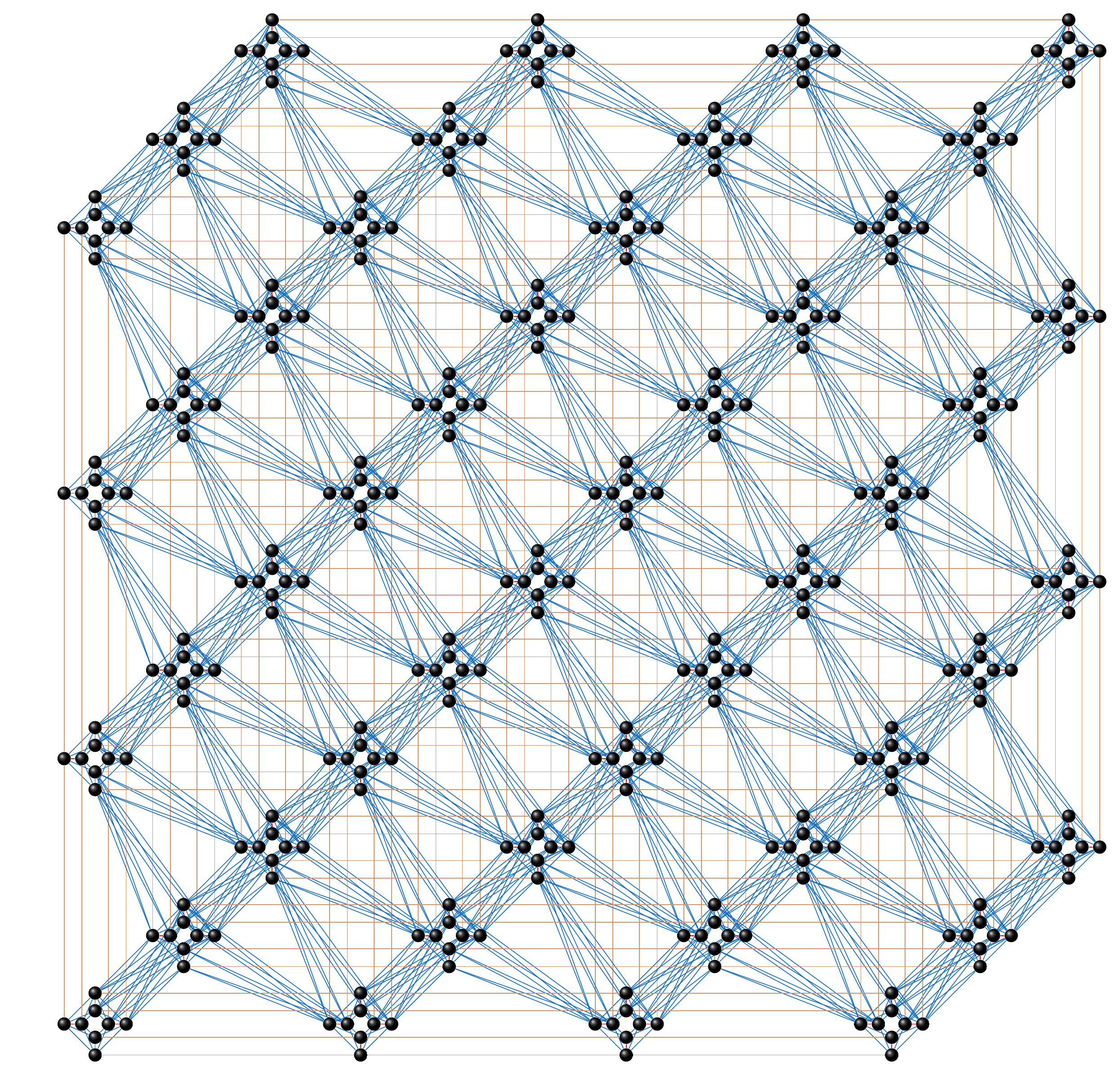}
         \caption{Pegasus}
         \label{fig:pegasus-qpu}
     \end{subfigure}
     \hfill
     \begin{subfigure}[b]{0.3\textwidth}
        \centering
        \includegraphics[width=\textwidth]{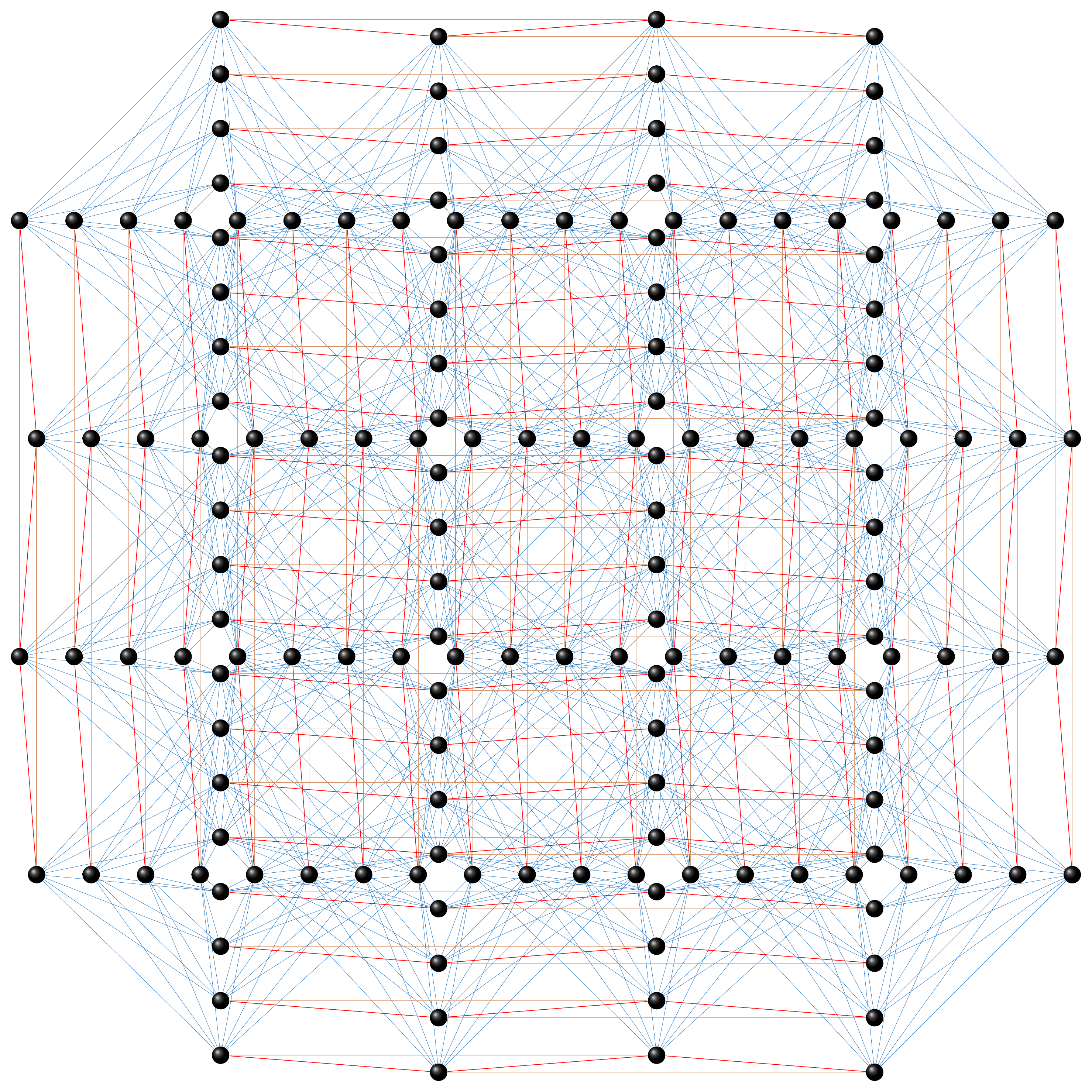}
         \caption{Zephyr}
         \label{fig:zephyr-qpu}
     \end{subfigure}
     \hfill
    \caption{Problems defined on Chimera, Pegasus, and Zephyr
    graphs~\cite{Boothby2021,Boothby2020,Dattani2019,Lanting2014}, employed in
    the past and current D-Wave quantum annealers, can be mapped to the Potts
    Hamiltonian on king's graph upon grouping $8$, $24$, and $16$ spins,
    respectively. Due to the large unit cell size of the latter two graphs, they
    require further processing. In particular, sparse connectivity structures
    between unit cells and GPU acceleration. Both are supported by
    \texttt{SpinGlassPEPS.jl} package.}
    \label{qpu}
\end{figure}

Our package can target the graphs that can be manufactured by quantum annealing
vendors (cf., D-Wave processors depicted in Fig.~\ref{qpu}). We provide an
extensive benchmark of our approach for those geometries elsewhere~\cite{main}.
   
\subsection{Solving Potts model - inpainting problem}
\label{MRFS}

\begin{figure}[h!]
        \begin{subfigure}[b]{0.3\textwidth}
        \centering
        \includegraphics[width=\textwidth]{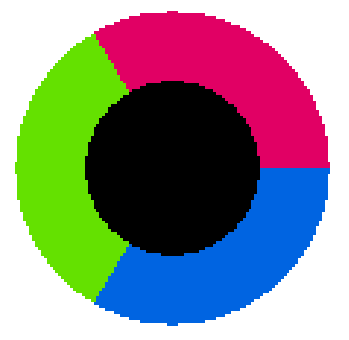}
         \caption{Input data}
    \end{subfigure}
    \hfill
    \begin{subfigure}[b]{0.3\textwidth}
        \centering
        \includegraphics[width=0.98\textwidth]{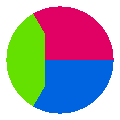}
         \caption{Best Found with PEPS}
      \end{subfigure}
     \hfill
     \begin{subfigure}[b]{0.29\textwidth}
        \centering
        \includegraphics[width=0.98\textwidth]{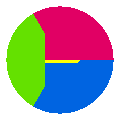}
         \caption{Droplet}
    \end{subfigure}
     \hfill
    \caption{ The picture in panel (a) shows the used inpainting problem. Data
    is given on the circular boundary, and the solution should fill in the black
    region. It is part of the OpenGM2 Benchmark dataset~\cite{openGM}. The
    picture in panel (b) shows the ground state obtained by
    \texttt{SpinGlassPEPS.jl}. Due to the introduced anisotropy by the grid used
    in discretization, the results show a bias toward axis-parallel
    edges~\cite{Lellmann2011}. Yellow region in (c) shows a low-energy
    excitation, \emph{i.e.}, a group of variables that should be collectively
    filliped to obtain another low-energy solution.}
        \label{fig:inpainting}
     \end{figure}

The Potts model defined in Eq.~(\ref{eq:Potts}) is general enough to describe a
wide array of phenomena. One of them is the problem of inpainting in computer
vision. It refers to filling in missing or corrupted parts of an image in a way
that makes the reconstruction visually plausible~\cite{Zeng2020}. In this
example, we use a relatively small benchmarking instance given
by~\cite{Lellmann2011, openGM}. It is a discretized triple junction inpainting
problem, as shown in Fig.~\ref{fig:inpainting}. The software accepts instances
formatted as in OpenGM Benchmark dataset~\cite{openGM} with only the nearest
neighbor and diagonal interactions. 

    \begin{minted}{Julia}
instance = "$(@__DIR__)/instances/triplepoint4-plain-ring.h5"
# We add size of picture in pixels
potts_h = potts_hamiltonian(instance, 120, 120)
    \end{minted}

In this case, when searching for droplets, one should set \texttt{mode}
parameter in \texttt{SingleLayerDroplets} to \texttt{:RMF}. 
    
    \begin{minted}{Julia}
droplets = SingleLayerDroplets(; max_energy = 100, min_size = 100, 
           metric = :hamming, mode=:RMF)
    \end{minted}

The remaining setup is analogous to the king’s graph scenario described in the
preceding section.
    
\section{Impact and Conclusion}

Tensor networks offer a powerful suite of tools, most notably employed in
quantum many-body simulations. Our package reduces the hurdles for applying
these methods to classical optimization~\cite{Cichocki2014}. Building on the
framework introduced in Ref.~\cite{Rams2021} -- originally targeting simpler
Chimera-like graphs -- \texttt{SpinGlassPEPS.jl} provides a flexible software
platform for a broader range of problem topologies, utilizing sparse tensor
structures, efficient use of GPU, etc.

As such, \texttt{SpinGlassPEPS.jl} acts as a practical reference tool to develop
and work with quantum and classical annealing technologies. The examples
presented show that the tensor network approach can achieve high-quality
solutions comparable to those of CPLEX, a state-of-the-art solver in operational
research~\cite{CPLEX}. Notably, for certain king's graphs~\cite{Chang2013}, our
solutions surpass those from Toshiba's Simulated Bifurcation Machines
(SBM~\cite{GotoSBM}), which is an unexpected result that merits further
investigation. This highlights a potential weakness in the SBM family of
algorithms, demonstrating the practical value of our software.
    
\section*{Acknowledgments}
This project was supported by the National Science Center (NCN), Poland, under
Projects: Sonata Bis 10, No. 2020/38/E/ST3/00269 (T.S., Z.M.) and
2020/38/E/ST3/00150 (A.D., M.R.) and Foundation for Polish Science (grant no
POIR.04.04.00-00-14DE/ 18-00 carried out within the Team-Net program co-financed
by the European Union under the European Regional Development Fund) (B.G.,
L.P.).

\bibliography{bibliography}
\end{document}